\documentclass[a4paper,twocolumn,11pt,accepted=2017-05-09]{quantumarticle}
\pdfoutput=1
\usepackage[utf8]{inputenc}
\usepackage[english]{babel}
\usepackage[T1]{fontenc}
\usepackage{hyperref}
\usepackage[normalem]{ulem}

\usepackage{quantikz}
\usepackage{braket}
\usepackage{verbatim}
\usepackage{amsfonts}
\usepackage{amssymb}
\usepackage{xcolor}
\usepackage{graphicx}
\usepackage{bm}
\usepackage{booktabs}
\usepackage[numbers,sort&compress]{natbib}
\usetikzlibrary{arrows.meta, positioning}
\usepackage{verbatim}
\usepackage{etoolbox}
\usepackage{graphicx}
\usepackage{subcaption} 
\usepackage{adjustbox}
\usepackage[ruled,vlined,linesnumbered]{algorithm2e}

\SetKwInOut{Input}{Input}
\SetKwInOut{Output}{Output}

\makeatletter
\providecommand\@afterenddocumenthook{}%
\makeatother

\usepackage{tikz}

\usepackage{lipsum}

\begin{document}

\title{Hybrid Quantum State Preparation via Data Compression}

\author{Emad Rezaei Fard Boosari}
\email{emad.boosari@gmail.com}
\orcid{0000-0002-4774-3915}
\author{Maryam Afsary}
\affiliation{Institute of Informatics, Faculty of Mathematics, Informatics and Mechanics,
University of Warsaw, Banacha 2c, 02–097 Warsaw, Poland}

\maketitle
\begin{abstract}
Quantum state preparation (QSP) for a general $n$-qubit state requires $\mathcal{O}(2^n)$ CNOT gates and circuit depth, making exact amplitude encoding (EAE) impractical for near-term quantum hardware.
We introduce an ancilla-free hybrid classical–quantum strategy that reduces this cost to $\mathcal{O}(\mathrm{poly}(n))$ for a broad class of compressible data. The method first applies a classical compression step to obtain a $d$-sparse representation of the input, loads this sparse vector using a sparse-state preparation routine, and then reconstructs the target state through a polynomial-depth quantum inverse transform. 
We evaluate the framework on synthetic benchmark signals and real biomedical time series using Fourier and Haar transforms, demonstrating 
substantial reductions in CNOT counts and circuit depth compared to EAE, together with competitive performance relative to the Fourier Series Loader (FSL).
The quantum simulation results show that combining classical data compression with quantum decompression provides a scalable framework for efficient QSP, reducing quantum overhead without requiring variational training or ancillary registers.
\end{abstract}

\section{Introduction}
Quantum state preparation, the coherent loading of classical data into quantum amplitudes, is an essential subroutine for quantum machine learning~\cite{biamonte2017, rebentrost2014}, Monte Carlo methods~\cite{montanaro2015quantum, stamatopoulos2020option}, Hamiltonian simulation~\cite{low2019hamiltonian}, linear-system solvers~\cite{harrow2009quantum}, and quantum differential-equation solvers~\cite{berry2014high}.  
However, preparing a generic $n$-qubit state requires $\mathcal{O}(2^{n})$ CNOT gates~\cite{mottonen2004, iten2016}, far beyond the limits of noisy intermediate-scale quantum (NISQ) hardware.

To mitigate this exponential overhead, several strategies have been proposed.  
Ancilla-assisted schemes~\cite{zhao2019state, zhang2024parallel} reduce depth at the cost of higher circuit width.  
Variational approaches~\cite{zoufal2019quantum} approximate states through training but can suffer from barren plateaus~\cite{mcclean2018barren}.  
Classical complexity-reduction frameworks offer related ideas as well, where pre-processing steps can suppress exponential growth (e.g., stratified multi-objective decision models that reduce $\mathcal{O}(2^n)$ scaling to $\mathcal{O}(n)$~\cite{sharifi2025stratified}).
Structure-exploiting methods, including sparse quantum state preparation (SQSP)~\cite{gleinig2021efficient, farias2025quantum, li2024nearly}, matrix product states (MPS)-based encodings~\cite{holmes2020efficient}, and truncated-series loaders~\cite{moosa2023linear, zylberman2024efficient} achieve polynomial costs. However, the performance of these methods can often degrade for real, noisy, non-smooth, or non-stationary datasets.
In addition, block-encoding–based approaches~\cite{mcardle2022quantum, o2025quantum, rosenkranz2025quantum} employ the quantum singular value transformation (QSVT) to approximate the target function. But the techniques are designed for fault-tolerant machines and rely on amplitude amplification, limiting their applicability to NISQ devices.

In this work, we introduce a general-purpose, ancilla-free hybrid classical–quantum algorithm for preparing compressible but non-sparse data. The key idea is to apply a classical transform–based compression step to obtain a $d$-sparse representation of the input, prepare this sparse vector using an SQSP technique, and then apply a polynomial-depth quantum inverse transform to reconstruct the target state. This replaces the exponential cost of EAE with a classical sparsification stage followed by an efficient quantum decompression stage.
Unlike variational or series-expansion approaches, the method is transform-agnostic, requires no ancillary qubits, and applies to data that may be non-smooth, noisy, or non-stationary, provided that it is classically compressible.

The framework applies to any classical data type for which a computationally efficient compression transform is available and whose action can be represented by a unitary operator, allowing the compressed coefficients to be recovered quantum mechanically. To illustrate the generality of this approach, we evaluate it on a range of representative datasets, including structured synthetic signals (periodic, sinc, and Gaussian waveforms) and real biomedical time series~\cite{pimentel2016toward}. Periodic signals are compressed using the discrete Fourier transform (DFT), while piecewise-smooth or non-stationary signals are handled via the discrete Haar wavelet transform (DHWT), which also admits efficient quantum implementations~\cite{li2018multi}.

To assess the quantum feasibility of the hybrid framework, we perform gate-level simulations of all hybrid circuits using Qiskit~\cite{javadi2024quantum} and Qibo~\cite{qibo_paper}. These simulations provide a detailed estimation of reconstruction fidelity and two important measures in circuit complexity: CNOT gate count and circuit depth for each benchmark signal. We compare these resource estimates against two baselines: (i) EAE~\cite{iten2016}, which serves as an upper bound on quantum cost, and (ii) state-of-the-art ancilla-free approximate QSP methods such as the Fourier series loader (FSL)~\cite{moosa2023linear}. 
Across all cases, we observe that classical sparsification translates directly into substantial reductions in quantum resources while maintaining high fidelity. 

The remainder of the paper is organized as follows. 
Section~\ref{sec:algorithm} describes the hybrid QSP framework in detail.
Section~\ref{sec:results} presents numerical simulations and resource estimates. 
Section~\ref{sec:discussion} compares the method with existing QSP techniques. 
Section~\ref{sec:conclusion} concludes with a summary and future directions.

\section{Hybrid QSP Algorithm}\label{sec:algorithm}
We now present the hybrid QSP algorithm, designed to reduce the cost of loading high-dimensional, compressible classical data into a quantum register. Consider an $N$-dimensional normalized vector
\begin{equation}\label{eq:input_x}
\bm{x} = [x_0, x_1, \dots, x_{N-1}]^T, 
\qquad 
\bm{x}^\dagger \bm{x} = 1.
\end{equation}
One can prepare this vector exactly as a quantum state by applying an operator $\mathcal{U}_{\text{exact}}$ on an $n$-qubit register:
\begin{equation}\label{eq:Psi}
\ket{\Psi} = \mathcal{U}_{\text{exact}} \ket{0}^{\otimes n} =\sum_{k=0}^{N-1} x_k \ket{k}, 
\qquad 
\braket{\Psi|\Psi} = 1,
\end{equation}
which requires $\mathcal{O}(2^n)$ CNOT gates and an exponentially deep circuit in $n$ ~\cite{iten2016}.
Our hybrid method reduces this cost, provided that
(i) a reversible classical compression algorithm exists for reducing the dimensionality of $\bm x$; and
(ii) a quantum implementation of the corresponding inverse transform (decompression) is available with lower computational complexity than $\mathcal{U}_{\text{exact}}$.

The degree of quantum resource reduction depends directly on the amount of classical data compression and the computational complexity of the decompression process. 
For the 1D signals studied in this work, the resulting sparsity is sufficient to reduce the overall cost to $\mathcal{O}(\mathrm{poly}(n))$.  
For higher-dimensional datasets such as images, the compressed representation is typically less sparse, and the hybrid method may offer only constant-factor reductions in the exponential overhead, without changing the asymptotic scaling.

The following sections detail the classical compression procedure and the corresponding quantum operations used for state reconstruction.

\subsection{Phase I: Classical Compression}
Phase~I performs classical pre-processing to produce a sparse representation of the input data that can later be loaded into a quantum register. The objective is to transform the dense vector $\bm{x}$ into a $d$-sparse ($d \ll N$) representation from which the original data can be reconstructed with minimal overhead.
To this end, one can follow the standard data compression workflow, which includes three steps:~\cite{sayood2017introduction, gersho2012vector}:  
\begin{itemize}
    \item \textbf{Transform}: apply a transformation to $\bm{x}$ to redistribute its energy into a smaller subset of coefficients;  
    \item \textbf{Coefficient reduction}: eliminate redundancy by pruning or reducing coefficient precision, typically via quantization, thresholding, or bit–allocation strategies;  
    \item \textbf{Entropy coding}: assign shorter codewords to frequent values, reducing storage or transmission costs.  
\end{itemize}
Since our hybrid algorithm aims to reconstruct compressed data within a quantum system, all classical steps must remain compatible with the laws of quantum mechanics.  
Hence, we avoid entropy coding and restrict our compression strategy to the following stages:  
(i) a unitary or orthogonal transformation, and  
(ii) an optional coefficient-reduction step.  

In principle, any coefficient-reduction mechanism implementable as a unitary operation could be incorporated into the framework. In this work, however, we adopt thresholding because of its simplicity and practical effectiveness. It is inherently lossy: once coefficients are discarded, no inverse quantum operation can recover them, and we need to renormalize the thresholded data.

Following the aforementioned strategy, we first apply a suitable unitary transform $\mathcal{U}_C$ to the input vector $\bm{x}$:
\begin{equation}\label{eq:compressed_X}
\bm{X} = \mathcal{U}_C \bm{x} = [X_0,\, X_1,\, \dots,\, X_{N-1}]^T.
\end{equation}
If $\bm{X}$ is exactly $d$–sparse, the compression is lossless, and no further processing is required.  
Otherwise, when $\bm{X}$ is not strictly sparse, we introduce a threshold parameter $\tau$ and discard all coefficients whose magnitude is below $\tau$, obtaining the truncated vector $\bm{X}(\tau)$.  
After truncation, the remaining coefficients must be renormalized to preserve unit norm:
\begin{equation}\label{eq:thresholding}
\bm{X}^r = \text{Normalize}\big(\bm{X}(\tau)\big).
\end{equation}
The normalized vector $\bm{X}^r$ is the final output of Phase I and serves as the input for the subsequent section.  

Classically, one would recover the approximate signal $\bm{x}(\tau)$ by applying the inverse transform $\mathcal{U}_C^{-1}$:
\begin{equation}\label{eq:reconstruction}
\bm{x}(\tau) = \mathcal{U}_C^{-1}\bm{X}^r = [\varphi_0,\, \varphi_1,\, \dots,\, \varphi_{N-1}]^T.
\end{equation}
In the next phase, this reconstruction is performed quantum mechanically by applying the quantum inverse transform 
$\mathcal{U}_Q^{-1}$, the quantum implementation of $\mathcal{U}_C^{-1}$.

\subsection{Phase II: Quantum Decompression}
Phase II performs the quantum reconstruction of the compressed data produced in Phase~I. This consists of two steps:
(i) preparing the sparse compressed vector $\bm{X}^r$ as a quantum state, and
(ii) applying the quantum inverse transform to recover the original state (exactly or approximately, depending on thresholding).

Since $\bm{X}^r$ is $d$-sparse, its preparation requires far fewer quantum resources than preparing the original dense vector.
One can prepare an $n$-qubit $d$-sparse vector efficiently by deploying any SQSP algorithm operator $\mathcal{P}$ \cite{gleinig2021efficient, malvetti2021quantum, ramacciotti2024simple, mozafari2022efficient, de2020circuit, de2022double, zhang2022quantum, tang2019quantum, li2024nearly, farias2025quantum} that maps the all-zero state to the corresponding quantum superposition:
\begin{equation}\label{eq:ket_phi}
\ket{\phi} = \mathcal{P}\ket{0}^{\otimes n} = \sum_{k=0}^{N-1} X_k^r \ket{k}.
\end{equation}
In this work, we adopt the construction of Farias et al~\cite{farias2025quantum}, which is efficient whenever the sparsity is polynomial in the number of qubits.

After preparing $\ket{\phi}$, the original state could be reconstructed by applying the quantum implementation of $\mathcal{U}_C^{-1}$:
\begin{equation}\label{eq:reconstructed_phi}
\ket{\Phi} = \mathcal{U}_Q^{-1}\ket{\phi} 
           = \sum_{k=0}^{N-1} \varphi_k \ket{k}.
\end{equation}
The operator $\mathcal{U}_Q^{-1}$ has been assumed to admit a polynomial–size quantum circuit, which holds for transforms such as the quantum Fourier transform (QFT), quantum discrete cosine transform, or quantum wavelet transforms. This ensures that the overall hybrid algorithm remains tractable, unlike EAE.  

A concise overview of the hybrid QSP workflow is provided in Algorithm~\ref{alg:hybrid_qsp}. The algorithm summarizes the classical pre-processing, SQSP, and inverse quantum transform required to realize the final state $\ket{\Phi}$.  

\begin{algorithm}[ht]
\SetAlgoLined
\Input{Normalized classical vector \( \bm{x} \in \mathbb{R}^N \) with \( \|\bm{x}\| = 1 \)}
\Output{Quantum state \( \ket{\Phi} = \sum_{k=0}^{N-1} \varphi_k \ket{k} \)}
\BlankLine
\SetKwBlock{CompressBlock}{\textbf{Classical compression:}}{end}
\CompressBlock{
  \textbf{Reversible transform:} \( \bm{X} \gets \mathcal{U}_C \bm{x} \).\\
  \SetKwBlock{ThreshBlock}{\textbf{Thresholding (optional):}}{end}
  \ThreshBlock{
    \small \( \bm{X}^r \gets \mathrm{Normalize}\big(\bm{X}(\tau)\big) \).\\
    \textit{If thresholding is skipped, set } \( \bm{X}^r \gets \bm{X} \). 
  }
}
\BlankLine
\textbf{SQSP:}\quad
\( \ket{\phi} = \mathcal{P}\ket{0}^{\otimes n} \).\\
\textbf{Quantum decompression:}\quad
\( \ket{\Phi} = \mathcal{U}_Q^{-1} \ket{\phi} \).
\caption{Hybrid QSP Algorithm.}
\label{alg:hybrid_qsp}
\end{algorithm}

Although our focus here is on amplitude encoding, the hybrid strategy readily 
generalizes to other embedding models. For example, the compressed coefficients 
may be encoded directly into computational-basis states,
\[
\ket{\phi} = \frac{1}{\sqrt{N}} \sum_{k=0}^{N-1} \ket{X_k^r}\ket{k},
\]
and subsequently decompressed using the corresponding reversible quantum decompression. Such hybrid constructions 
have been explored in image preparation using JPEG-style transforms~\cite{jiang2018novel}.  
This illustrates the flexibility of the hybrid QSP paradigm as a general 
blueprint for combining classical compression with quantum decompression across 
a variety of data types and encoding schemes.

\subsection{Computational Complexity and Metrics}
The overall cost of the hybrid QSP algorithm naturally separates into two parts: a classical pre–processing stage that compresses the data and a quantum stage that reconstructs the original signal.  

On the classical side, the dominant cost arises from applying the chosen transform $\mathcal{U}_C$ (e.g., Fourier, cosine, or wavelet family), together with coefficient selection or thresholding. Since $\mathcal{U}_C$ is typically realizable in quasi-linear time, the compression stage scales as $\mathcal{O}(N \log N)$ for an input of length $N$, while thresholding and normalization are linear in $N$ and therefore do not affect the overall asymptotic order.

In practice, selecting the compression parameters, such as the choice of transform, decomposition depth, or threshold $\tau$, may introduce an additional classical overhead.
For many data modalities, however, classical signal-processing theory already provides well-established choices (e.g., Fourier for periodic signals, Haar or wavelets for piecewise-smooth or non-stationary signals), so only minimal tuning is required.
If parameter selection consists of evaluating a small number of candidate configurations or optimizing a fidelity-based cost function, the total complexity remains polynomial, typically $\mathcal{O}(K N \log N)$, where $K$ is the number of evaluations.
More adaptive or data-driven tuning procedures may increase pre-processing cost, but they do not affect the quantum complexity of the hybrid method.

On the quantum side, the first computational cost corresponds to encoding the $n$-qubit $d$-sparse vector $\bm{X}^r$.  
The required circuit size (the number of elementary quantum gates) depends on the chosen SQSP methods.  
For example, the algorithm of Li~\emph{et al.}~\cite{li2024nearly} prepares the state $\ket{\phi}$ with a circuit of size $\mathcal{O}\!\left(\frac{nd}{\log n}+n\right)$, while the approach of Farias~\emph{et al.}~\cite{farias2025quantum} requires $\mathcal{O}(\mathrm{poly}(n))$ CNOT gates.  
Thus, for a $d$-sparse state, the overall gate complexity of this step scales polynomially in $n$. 

The second contribution is the implementation of the inverse transform $\mathcal{U}_Q^{-1}$. Its cost depends on the chosen transform, but under the assumption that $\mathcal{U}_Q^{-1}$ admits a polynomial gate complexity, the overall hybrid algorithm requires only polynomial resources in $n$.

Taken together, both the SQSP preparation and quantum decompression stages require only $\mathcal{O}(\mathrm{poly}(n))$ gates.  
However, since different compression methods lead to different circuit structures, the dominant complexity term cannot generally be specified.  
In some cases, the SQSP stage may dominate the overall cost, while the decompression circuit contributes more significantly in others.

\subsubsection{Metrics}
To evaluate performance, we rely on four complementary metrics. The first is the \emph{compression ratio} (CR), defined as
\begin{equation}
\mathrm{CR} = \frac{N}{d},
\end{equation}
which quantifies how effectively the classical compression step compacts information. Larger CR values correspond to stronger compression and, in principle, greater savings in quantum resources.  

The second metric is the number of CNOT gates needed to implement the quantum phase, which reflects the practical feasibility of running the algorithm on near-term quantum hardware. The third metric is circuit depth, which refers to the minimum number of time steps (or layers) required to execute a quantum circuit, assuming that gates acting on disjoint sets of qubits can be applied in parallel. In other words, the depth is the length of the longest path through the circuit when parallel execution is allowed  \cite{nielsen2010quantum,moore2001parallel}.

Finally, the reconstruction accuracy between the target state $\ket{\Psi}$ and the prepared state $\ket{\Phi}$ is quantified by the \emph{trace distance} (TD), defined for pure states as
\begin{equation}
D_{\mathrm{tr}}(\Psi,\Phi) = \sqrt{1 - \mathcal{F}}.
\end{equation}
where $\mathcal{F} = |\braket{\Psi|\Phi}|^2$ is the state fidelity. For identical states, $D_{\mathrm{tr}} = 0$ and $D_{\mathrm{tr}} = 1$ corresponds to orthogonal ones.  
We define an admissible tolerance $\epsilon$ such that
\begin{equation}
D_{\mathrm{tr}}(\Psi,\Phi) < \epsilon,
\end{equation}
ensuring that the reconstructed quantum state remains within the desired fidelity range of the target state.

Together, these four metrics CR, CNOT count, circuit depth, and TD provide a comprehensive framework for evaluating the trade–off between classical compressibility and quantum resource efficiency across different transforms and operating regimes.

\section{Simulation Results}\label{sec:results}
In this section, we present simulation results of the proposed hybrid algorithm for different input data to demonstrate its practical effectiveness in loading data into a quantum system. The results are grouped into two categories:
\begin{itemize}
\item \textbf{Lossless Hybrid Preparation}: Signals that are intrinsically sparse after transformation, allowing unit fidelity and exponentially fewer gates than EAE.
\item \textbf{Approximate Hybrid Preparation}: Signals that need thresholding to reach a sparse representation, enabling high compression with controllable error.
\end{itemize}

All quantum simulations have been carried out by {Qiskit}~\cite{javadi2024quantum} and {Qibo}~\cite{qibo_paper}, which have been employed for circuit synthesis and resource estimation.

\subsection{Lossless Hybrid Preparation}\label{sec:perfect_qsp}
We begin with two illustrative cases where our protocol reconstructs the input signals precisely, achieving unit fidelity and zero approximation error ($\epsilon = 0$). 
These lossless examples are useful benchmarks: they achieve exponential savings in gate complexity relative to EAE, while preserving perfect fidelity ($D_{\text{tr}}=0$).

\subsubsection{Periodic Signal}\label{sec:periodic_signal}
\begin{figure}[t]
  \centering
  \begin{subfigure}[b]{0.48\textwidth}
    \includegraphics[width=\textwidth]{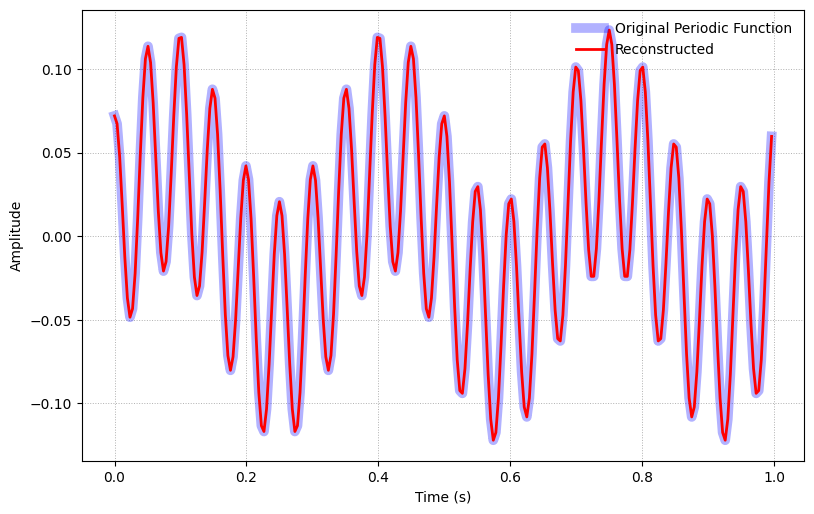}
    \caption{}
    \label{fig:multi-periodic-signal}
  \end{subfigure}  
  \hfill
  \begin{subfigure}[b]{0.48\textwidth}
    \includegraphics[width=\textwidth]{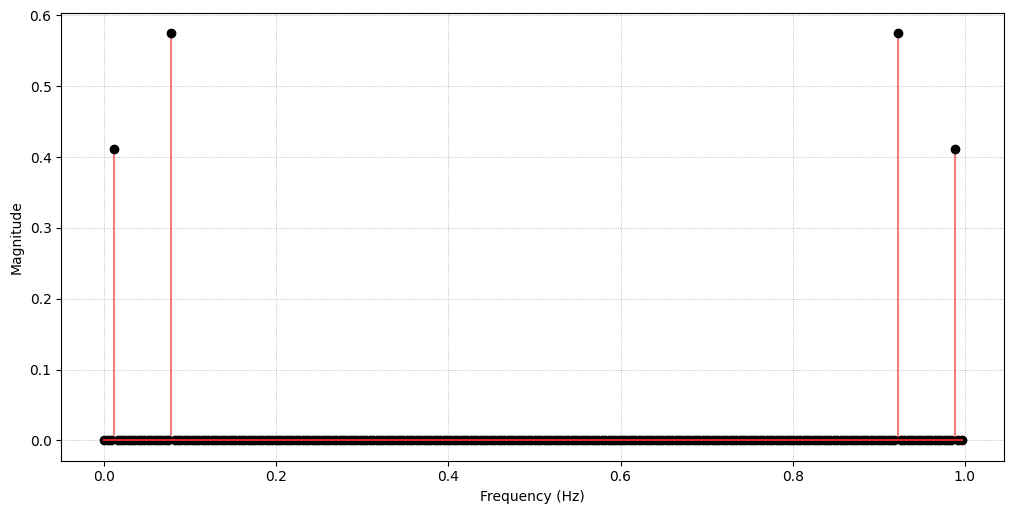}    
    \caption{}
    \label{fig:DFT_multi_periodic}
  \end{subfigure}  
  \caption{(a) Original multi-frequency periodic signal (light blue) and its exact reconstruction (red) via hybrid QSP. (b) Frequency-domain representation obtained via DFT, revealing four non-zero coefficients.}
    \label{fig:periodic_signail_&_DFT}
\end{figure}
Periodic signals provide a natural testbed for our framework because they are intrinsically sparse in the Fourier domain: only a few frequency components carry non-zero amplitude, while all others vanish exactly. Mathematically, the DFT of a length-$N$ signal is defined as
\begin{equation}\label{eq:DFT_equation}
    X_k = \frac{1}{\sqrt{N}} \sum_{j=0}^{N-1} x_j \, e^{- 2 \pi i k j / N},     
\end{equation}
for $k \in \{0, 1, \dots, N-1\}$.

\begin{figure}[ht] 
    \centering 
    \includegraphics[width=0.95\linewidth]{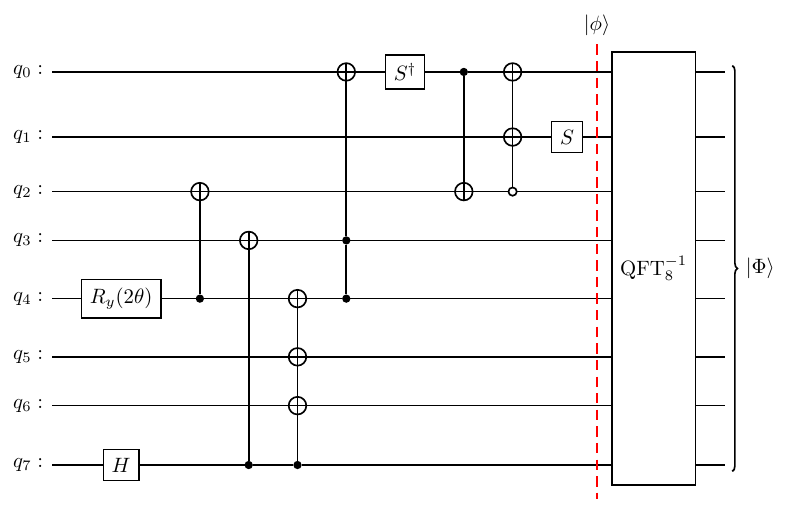} 
    \caption{Hybrid QSP circuit for the multi-frequency periodic signal in Fig.~\ref{fig:multi-periodic-signal}.  
    The 4-sparse frequency-domain state is initialized using a single $R_y(2\theta)$ rotation with $\theta \approx 0.9505$ radians and prepared with only 14 CNOT gates and 15 single-qubit rotations. An inverse QFT is then applied to recover the exact original time-domain signal. For resource estimation, each Toffoli gate could be decomposed into 6 CNOTs and 9 single-qubit gates \cite{nielsen2010quantum}.}    
    \label{fig:QFT_inverse_circuit} 
\end{figure} 
Figure~\ref{fig:multi-periodic-signal} shows a multi-periodic signal of length $N=256$ together with its frequency spectrum in Fig.~\ref{fig:DFT_multi_periodic}.  
Only four Fourier coefficients are non-zero:  
\begin{eqnarray}
    \begin{cases}
        X_{3}   = iA, \quad X_{253} = -iA,\\
        X_{20}  = X_{236} = B,        
    \end{cases}
\end{eqnarray}
with $A = 0.41099$ and $B = 0.57539$. The transformed state is therefore genuinely sparse, yielding a $\mathrm{CR}=64{:}1$.  
In this case, an 8-qubit 4-sparse state $\ket{\phi}$ can be prepared efficiently via any proper SQSP algorithm using an operator $\mathcal{P}$:  
\begin{equation}
    \ket{\phi} = \mathcal{P}\ket{0}^{\otimes 8}.
\end{equation}
The original signal is then recovered by applying the inverse QFT, which has complexity $\mathcal{O}(n^2)$~\cite{shor1994algorithms, nielsen2010quantum}:
\begin{eqnarray}
    \ket{\Phi} = \text{QFT}^{-1}_n \ket{\phi} = \sum_{\ell=0}^{N-1} \varphi_\ell \ket{\ell}.
\end{eqnarray}

The full hybrid circuit (SQSP $+$ inverse QFT) is shown in Fig.~\ref{fig:QFT_inverse_circuit}. Its Qiskit synthesis required 82 CNOT and 107 single-qubit gates, while the circuit depth is 74. 
This represents nearly a $\sim 3\times$ saving in entangling gates and a $\sim 7\times$ in circuit depth, in comparison with EAE \cite{iten2016}.

The resulting state is shown in Fig.~\ref{fig:multi-periodic-signal}, where the red curve (hybrid QSP) overlaps perfectly with the original light blue signal, confirming unit fidelity. 
This example highlights a central feature of the hybrid protocol: whenever a signal admits an exactly sparse representation under a reversible classical transform, perfect QSP can be achieved with substantially fewer gates and dramatically shallower circuits, a decisive advantage for NISQ devices.

\subsubsection{Piecewise-constant Signal}\label{sec:step_signal}
Beyond the periodic signals that our algorithm can prepare flawlessly, another example could be a piecewise-smooth function. It exhibits localized features such as discontinuities, constant plateaus, or gradual transitions, making it highly compressible via the DHWT~\cite{mallat1999wavelet}.  
Unlike the DFT, which provides the most effective global frequency representation for stationary signals, wavelet transforms offer a multi-resolution analysis better adapted to non-stationary features. As a result, functions such as piecewise-constant signals can be represented compactly with only a handful of non-zero wavelet coefficients.
\begin{figure}[t]
    \centering
    \includegraphics[width=0.48\textwidth]{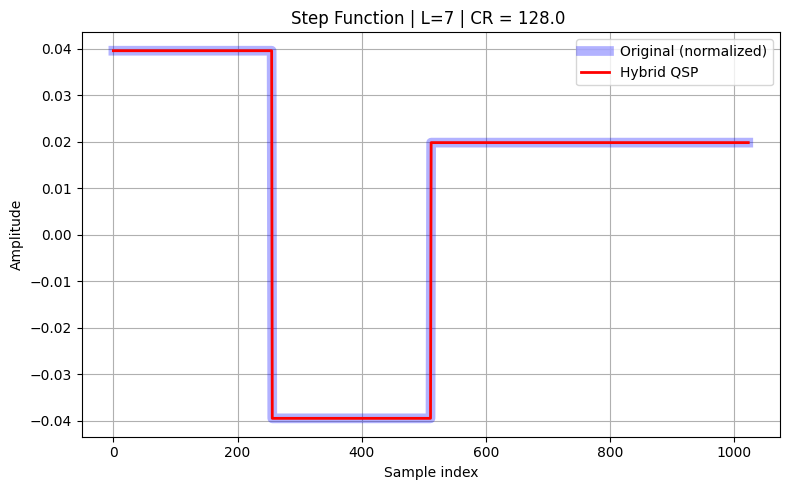}
    \caption{Illustrative representation of the original piecewise-constant signal (light blue) versus its quantum preparation using the proposed hybrid algorithm (red).}
    \label{fig:step_function_original}    
\end{figure}

\begin{figure*}
    \centering
    \begin{subfigure}[b]{0.48\textwidth}
        \centering
        \includegraphics[width=\linewidth]{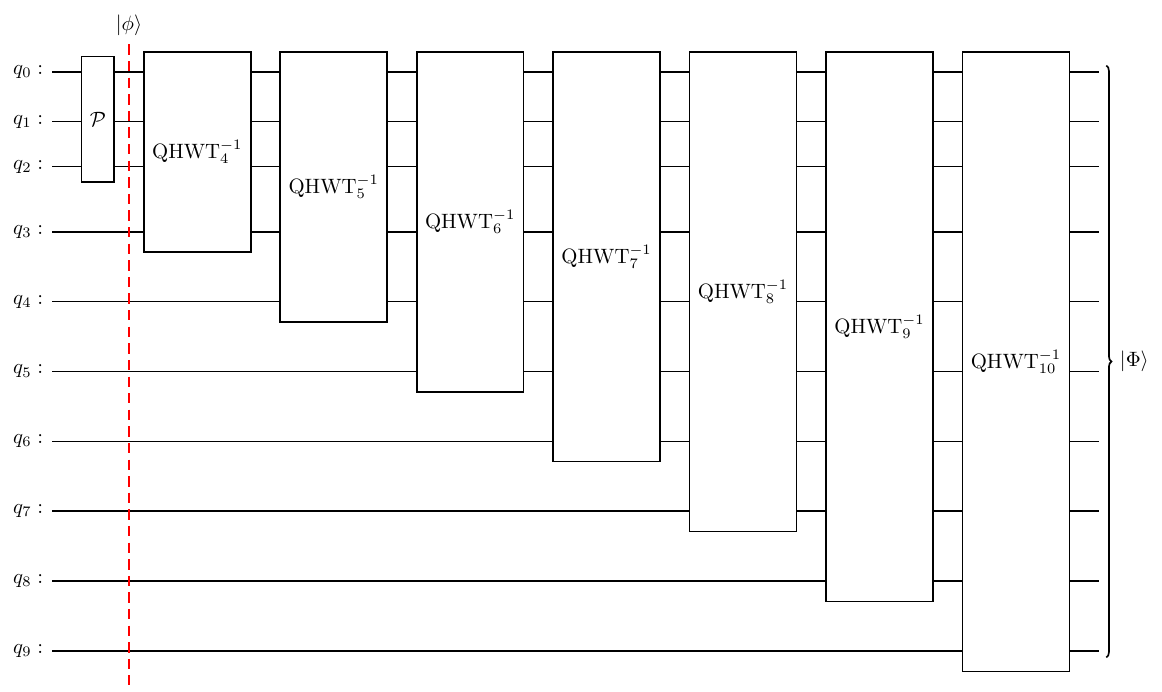}
        \caption{}
        \label{fig:hybrid_qsp__inverse_circuit}
    \end{subfigure}
    \hfill
    \begin{subfigure}[b]{0.48\textwidth}
        \centering
        \includegraphics[width=\linewidth]{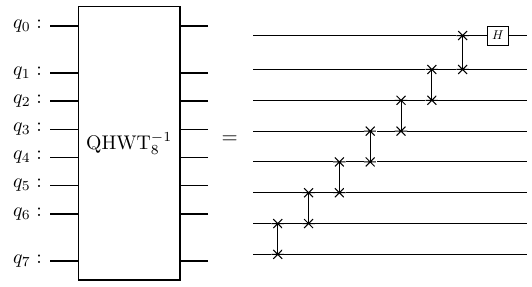}
        \caption{}
        \label{fig:qhwt_single_circuit}
    \end{subfigure}
    \caption{Hybrid QSP for the piecewise-constant signal. 
    (a) The quantum circuit for the hybrid QSP, including the sparse preparation operator $\mathcal{P}$ and a 7-level inverse packet QHWT for data decompression.
    (b) Expanded view of $\text{QHWT}_8^{-1}$ block.}
    \label{fig:piecewise_qhwt}
\end{figure*}

In general, a discrete wavelet transformation (DWT) decomposes a signal $\bm{x}$ into low-frequency (approximation) and high-frequency (detail) components by filtering and down-sampling. Formally, this can be written as
\begin{equation}
\begin{cases}
A_k = \sum_{j} x_j \cdot h_{2k - j}, \\
D_k = \sum_{j} x_j \cdot g_{2k - j},    
\end{cases}
\end{equation}
where $h_k$ and $g_k$ denote the low-pass and high-pass filters, respectively. Down-sampling by two ensures that the transformed signal retains the same dimension.  

It is important to note that two principal forms of the DWT exist.  The more common, \emph{pyramid form} (or multi-resolution) recursively decomposes only the approximation coefficients ($A_k$) at each level. This structure is computationally efficient in the classical setting ($\mathcal{O}(2^n)$ complexity) but challenging to realize in a quantum circuit \cite{li2019quantum}.  
By contrast, the \emph{packet form} recursively decomposes both the approximation ($A_k$) and detail coefficients ($D_k$), resulting in richer hierarchical information. While packet wavelets incur higher classical complexity ($\mathcal{O}(n2^n)$), they admit more straightforward quantum implementations with polynomial gate counts ($\mathcal{O}(n^2)$) and therefore provide a more suitable foundation for hybrid QSP~\cite{li2018multi}.  
For this reason, we adopt the packet-wavelet structure throughout this paper.

For piecewise-constant signals, the DHWT is especially effective, since its piecewise-constant basis functions align naturally with the discontinuities of the input. The Haar filters are given by
\begin{equation}
\bm{h} = \left(\tfrac{1}{\sqrt{2}},\, \tfrac{1}{\sqrt{2}}\right), \qquad
\bm{g} = \left(\tfrac{1}{\sqrt{2}},\, -\tfrac{1}{\sqrt{2}}\right).
\end{equation}
For illustration, the single-level Haar analysis operator on a length-8 signal acts on neighboring pairs as
\begin{equation}
\small
\begin{pmatrix}
h_0 & h_1 & 0   & 0   & 0   & 0   & 0   & 0 \\
0   & 0   & h_0 & h_1 & 0   & 0   & 0   & 0 \\
0   & 0   & 0   & 0   & h_0 & h_1 & 0   & 0 \\
0   & 0   & 0   & 0   & 0   & 0   & h_0 & h_1 \\
g_0 & g_1 & 0   & 0   & 0   & 0   & 0   & 0 \\
0   & 0   & g_0 & g_1 & 0   & 0   & 0   & 0 \\
0   & 0   & 0   & 0   & g_0 & g_1 & 0   & 0 \\
0   & 0   & 0   & 0   & 0   & 0   & g_0 & g_1
\end{pmatrix}.
\end{equation}

To test this in practice, we considered a piecewise-constant signal of length $N=1024$, shown in Fig.~\ref{fig:step_function_original}.  
A 7-level packet DHWT yields an 8-sparse coefficient vector, giving a $\mathrm{CR}=128{:}1$ without thresholding.
Because the packet DHWT places all non-zero coefficients in the first eight positions, the compressed vector occupies only the three least significant basis states.
Hence, instead of preparing a full 10-qubit 8-sparse state, we prepare only the lower three qubits:
\begin{equation}
\ket{\phi} \;=\; \ket{0}^{\otimes 7} \otimes \mathcal{P}\ket{000}
= \ket{0}^{\otimes 7} \otimes \sum_{k=0}^7 X^r_k \ket{k},
\end{equation}
which can be encoded using just 4 CNOT and 7 single-qubit gates.

The final step applies the inverse quantum Haar wavelet transform (QHWT) to the compressed state $\ket{\phi}$. 
The reconstruction applies a 7-level inverse QHWT. Each $n$-qubit block $\text{QHWT}^{-1}_n$ consists of one Hadamard gate and $(n-1)$ SWAPs, yielding a depth of $3n-2$.
Figure~\ref{fig:hybrid_qsp__inverse_circuit} shows the full 7-level inverse transform, while Fig.~\ref{fig:qhwt_single_circuit} provides the structure of a single $\text{QHWT}^{-1}_8$ block as a representative example.
Since the pattern of the quantum reconstruction circuit repeats across all subsequent case studies, we omit redundant circuit diagrams later in the paper.

Overall, the quantum decompression requires 126 CNOT gates and a depth of 46, achieving an $\sim8\times$ reduction in entangling gates and a $\sim45\times$ reduction in depth compared with EAE—while maintaining unit fidelity.
This example demonstrates the optimal, lossless performance of the hybrid approach when the transform-domain representation is exactly sparse.

Finally, note that the examples in this section correspond to special cases where the transformed coefficients vanish exactly, enabling perfect reconstruction.
In the next section, we consider more realistic signals for which sparsity must be induced through thresholding, introducing a tunable trade-off between compression and fidelity.

\subsection{Approximate Hybrid Preparation}\label{sec:approx_qsp}
\begin{table*}[t]
\centering
\caption[Hybrid QSP resource summary]{\textbf{Quantum resource summary for the hybrid QSP.}  
The table reports the CR, the number of CNOT gates and circuit depth for two components of the hybrid framework:  
(i) the SQSP stage implemented using the Farias~\emph{et~al.}~\cite{farias2025quantum} algorithm, and  
(ii) the quantum decompression stage based on the inverse QHWT.  
Each signal type is analyzed for the specified number of qubits ($n$) and decomposition level ($L$).  
The rightmost columns list the achieved reduction factors in CNOT count and circuit depth relative to EAE \cite{iten2016}, together with the final TD quantifying reconstruction accuracy.}
\label{tab:tab1}
\setlength{\tabcolsep}{4pt}
\renewcommand{\arraystretch}{1.1}
\begin{tabular}{l c c c cc cc cc c}
\toprule
Signal & $n$ & $L$ & CR
& \multicolumn{2}{c}{SQSP (Farias)}
& \multicolumn{2}{c}{Decompression}
& \multicolumn{2}{c}{Reduction vs.\ EAE}
& TD \\
\cmidrule(lr){5-6}\cmidrule(lr){7-8}\cmidrule(lr){9-10}
& & & & CNOTs & Depth & CNOTs & Depth & CNOT & Depth  & \\
\midrule
Sinc Function       & 15 & 10 & 298{:}1  &   494  &  760   &  285   &   70   &  $\sim42\times$  & $\sim79\times$  & 0.035 \\
Single Gaussian     & 15 & 13 & 745{:}1  &   210  &  470   &  312   &   79   &  $\sim63\times$  & $\sim119\times$ & 0.010 \\
Gaussian Mixture    & 15 & 12 & 328{:}1  &   416  &  677   &  306   &   76   &  $\sim45\times$  & $\sim87\times$  & 0.017 \\
PPG                 & 16 & 13 &  20{:}1  &   17k  &  25k   &  351   &   82   & $\sim4\times$    & $\sim6\times$   & 0.069 \\
\bottomrule
\end{tabular}

\vspace{2mm}
\end{table*}

In the preceding section, the idealized showcases demonstrated that our hybrid algorithm performs effectively for structured signals.  
To further assess its applicability, we now turn to more realistic signals that exhibit irregularity, noise, or sudden changes. 
In these settings, the DHWT remains highly effective, providing a reversible and 
multi-resolution representation for non-stationary signals.

In the following examples, each signal is transformed using the packet DHWT and then thresholded to enforce sparsity. This allows us to examine how the hybrid framework balances compression, circuit cost, and reconstruction fidelity in practical scenarios.

\subsubsection{The Sinc Signal}\label{sec:sinc_example}
\begin{figure}
  \centering
  \includegraphics[width=0.48\textwidth]{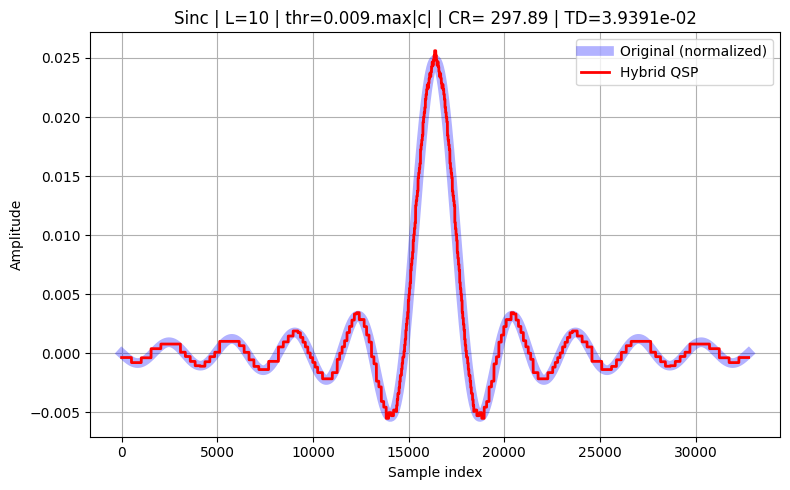}
  \caption{Approximate hybrid QSP of a sinc signal using the Haar wavelet basis.  
  The reconstructed signal (red) closely matches the original (light blue), 
  demonstrating strong fidelity after high compression.}
  \label{fig:sinc-haar}
\end{figure}

The sinc function serves as a canonical example of a smooth oscillatory waveform. Defined by
\begin{equation}
x(t) = \frac{\sin(\pi t)}{\pi t}, \quad t \in \mathbb{R},
\end{equation}
it decays rapidly away from the origin and is therefore highly compressible under DWTs.
We sampled the signal at $N = 2^{15}$ equispaced points, normalized it to unit $\ell_2$ norm, and applied a 10-level packet DHWT.

To enforce sparsity, coefficients below $0.9\%$ of the maximum magnitude were discarded, yielding a 110-sparse vector and a CR of 
$\approx 298{:}1$.
The cost of preparing the 15-qubit sparse state and performing quantum decompression is reported in Table~\ref{tab:tab1}.
Relative to EAE, the hybrid approach reduces CNOT count by $\sim42\times$ and circuit depth by $\sim79\times$, while achieving high reconstruction fidelity ($\mathcal{F}=99.930\%$). 
Fig.~\ref{fig:sinc-haar} shows that the hybrid reconstruction is nearly indistinguishable from the target signal.

\subsubsection{Gaussian Signal}\label{sec:gaussian_signal}
\begin{figure}
\centering 
\includegraphics[width=0.48\textwidth]{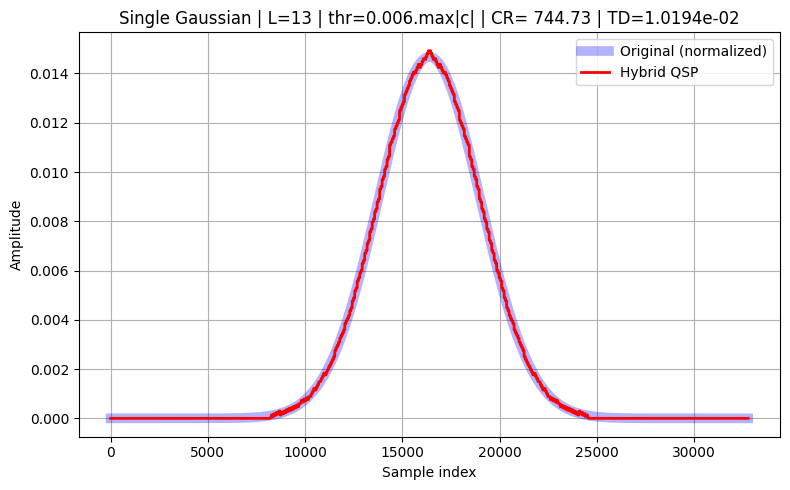} 
\caption{Approximate hybrid QSP of a Gaussian signal using the DHWT.
The reconstructed waveform (red) closely aligns with the original (light blue), indicating high fidelity.}
\label{fig:gaussian-single-haar} 
\end{figure}

Gaussian waveforms represent another important family of smooth, localized signals.
They appear in quantum chemistry, harmonic oscillator eigenstates, machine learning kernels, and many physical models~\cite{kitaev2008wavefunction, nielsen2010quantum, stamatopoulos2020option, mcclean2016theory}.
We begin with a single Gaussian waveform as a baseline and later extend the analysis to Gaussian mixtures, better representing structured real-world data.  

A one-dimensional Gaussian waveform is defined as
\begin{equation}
    g(x;\mu,\sigma) = \exp\!\left[-\frac{(x-\mu)^2}{2\sigma^2}\right],
    \label{eq:gaussian}
\end{equation}
where $\mu$ and $\sigma$ denote the mean and standard deviation, respectively.  
The discretized signal is normalized to unit $\ell_2$ norm to ensure compatibility with quantum state encoding.  
As a baseline test, we considered a Gaussian signal with parameters $(\mu = 0.0,\ \sigma = 0.8)$ and length $N = 2^{15}$, sampled uniformly over the interval $x \in [-5, 5]$ (Fig.~\ref{fig:gaussian-single-haar}).  

By applying a 13-level packet DHWT with a threshold set to $0.6\%$ of the maximum coefficient magnitude, the signal becomes strongly sparse, retaining only 44 coefficients ($\mathrm{CR}\approx 745{:}1$). 
The simulation results indicate nearly a $\sim63\times$ reduction in CNOT gates and a $\sim119\times$ reduction in circuit depth compared to EAE (see Table~\ref{tab:tab1}).  
Figure~\ref{fig:gaussian-single-haar} illustrates the reconstructed waveform obtained via the hybrid QSP approach, showing excellent agreement between the prepared and target signals with a high fidelity ($\mathcal{F} = 99.999\%$).  

\begin{figure}
\centering 
\includegraphics[width=0.48\textwidth]{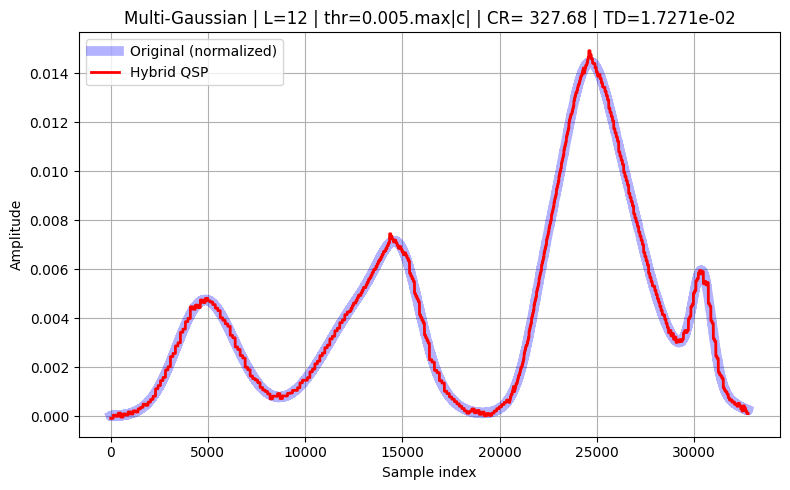} 
\caption{Approximate hybrid QSP of a Gaussian mixture signal using the packet DHWT. 
The reconstructed waveform (red) closely follows the original (light blue), confirming accurate recovery under compression. 
The signal consists of $K = 12$ Gaussian components with random centers $\mu_k \in [-4.5,4.5]$, widths $\sigma_k \in [0.12,0.60]$, and amplitudes $a_k \in [0.30,1.00]$. 
A small additive white noise term ($0.001$ standard deviation) is included to model fine-scale variations, 
and the resulting waveform is normalized to unit $\ell_2$ norm for QSP.}
\label{fig:gaussian-mixture-haar} 
\end{figure}

We next examined mixtures of Gaussians, which provide more complex, structured, and realistic signals with multiple peaks.
A mixture of $K$ Gaussians is defined as
\begin{equation}
    f(x) = \sum_{k=1}^{K} a_k \, g(x;\mu_k,\sigma_k),
    \label{eq:gaussian-mixture}
\end{equation}
where $a_k$ denotes the amplitude, and $\mu_k, \sigma_k$ specify the center and width of each component.  
For our simulations, we generated a mixture of $K=12$ Gaussians with random centers, widths, and amplitudes and a small white-noise term to introduce fine-scale variability (see Fig.~\ref{fig:gaussian-mixture-haar}). 
The signal length was again $N=2^{15}$, sampled uniformly over $x \in [-5, 5]$.  

Similar to the single-Gaussian case, the Gaussian mixture exhibits strong sparsity under the packet DHWT. Using a 12-level decomposition and with a threshold set to $0.5\%$ of the maximum coefficient magnitude, the signal compresses to 100 coefficients ($\mathrm{CR}\approx 328{:}1$). The corresponding quantum simulation shows reductions of $\sim45\times$ in CNOT count and $\sim87\times$ in circuit depth relative to EAE, while achieving high fidelity ($\mathcal{F}=99.985\%$), as reported in Table~\ref{tab:tab1}.

The Fig.~\ref{fig:gaussian-mixture-haar} confirms that the hybrid QSP efficiently processes complicated structured signals and preserves high fidelity even in overlapping spectral components.

\subsubsection{Application to Biomedical Signals: PPG Dataset}\label{sec:PPG}
\begin{figure}[t]
\centering 
\includegraphics[width=0.45\textwidth]{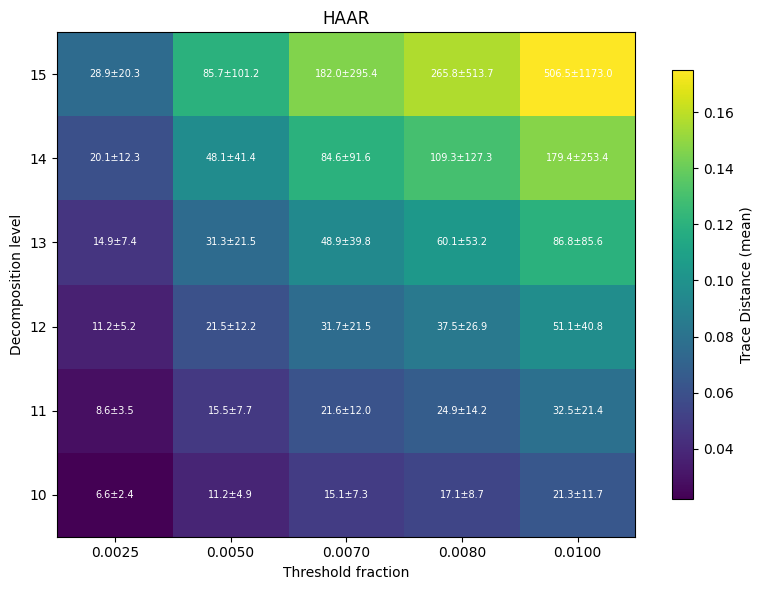} 
\caption{Heatmap of mean TD across 53 PPG recordings under DHWT compression.  
The background color encodes the mean $D_{\mathrm{tr}}$, while overlaid numbers report the mean CR with one standard deviation ($\pm$).  
The horizontal axis shows the threshold fraction ($\tau$) for coefficient pruning, and the vertical axis the decomposition level $L$.  
This representation highlights the regions where strong compression (high CR) can be achieved with minimal fidelity loss, guiding parameter selection for hybrid QSP of biomedical signals.}
\label{fig:PPG-dataset} 
\end{figure}

Biomedical time-series data, such as photoplethysmography (PPG), are high-dimensional, nonstationary, and heavily affected by noise and motion artifacts.
We analyzed 53 PPG recordings from the BIDMC PPG and Respiration dataset~\cite{pimentel2016toward}, available through PhysioNet~\cite{goldberger2000physiobank}.
Each waveform contains approximately $60{,}001$ samples, corresponding to $\lceil \log_2 N \rceil = 16$ qubits after zero-padding.
The EAE of such a state would require on the order of 60,000 CNOT gates and a circuit depth exceeding $10^5$, placing it far beyond the capabilities of near-term devices.
This makes PPG signals an ideal candidate for hybrid QSP, where classical compression can greatly reduce the quantum resources required for state preparation.

To identify suitable compression parameters, we evaluated each waveform across multiple decomposition levels $L$ and threshold values $\tau$. The Fig.~\ref{fig:PPG-dataset} shows the resulting mean TD (background color) and mean CR (overlaid values with standard deviation). 
Valid operating regimes for hybrid QSP correspond to regions where the number of retained coefficients remains polynomial in $n$:
\begin{equation}
    n^2\le d \le n^3 \Longleftrightarrow{}  15 \le \text{CR} \le 235.
\end{equation}
High CR values close to the $n^2$ regime tend to reduce fidelity, while lower CR values near $n^3$ increase the SQSP gate overhead. We therefore select an intermediate CR that balances reconstruction accuracy with quantum resource consumption.
\begin{figure}[t]
\centering 
\includegraphics[width=0.45\textwidth]{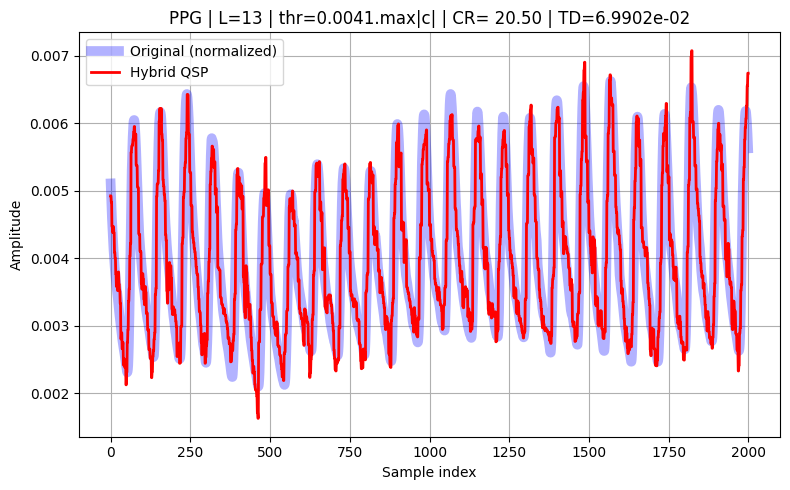} 
\caption{Hybrid QSP reconstruction of a representative PPG signal using the DHWT.  
The red curve shows the reconstructed waveform from the hybrid QSP algorithm, closely matching the original signal (light blue).}
\label{fig:PPG-haar} 
\end{figure}

As a representative case, we analyzed a single PPG signal from the dataset to illustrate the practical performance of our hybrid algorithm. 
Fig.~\ref{fig:PPG-haar} presents a 2000-sample segment of the PPG signal (light blue). 
Using a 13-level packet DHWT with threshold $\tau = 0.0041$, the signal compresses to 3197 coefficients ($\mathrm{CR} \approx 20$).
Detailed resource estimates in Table~\ref{tab:tab1} indicate that the Haar-based preparation produces circuits with approximately $\sim4\times$ fewer CNOT gates and $\sim 6\times$ shallower depth than EAE, while maintaining high fidelity ($\mathcal{F}= 99.755\%$). 
Although the reconstructed PPG signal (red curve in Fig.~\ref{fig:PPG-haar}) exhibits slightly lower fidelity than the synthetic benchmark signals, it remains within a practical and acceptable range for noisy biomedical data.

Taken together, the PPG study demonstrates that the proposed hybrid QSP framework extends beyond analytical benchmarks to realistic biomedical signals.  
By combining the Haar wavelet compression with SQSP, the method yields significant reductions in quantum resources while preserving reconstruction accuracy suitable for downstream quantum algorithms.

\section{Discussion}\label{sec:discussion}

\begin{table*}[t]
\centering
\caption{Quantum resource estimates for the FSL applied to the benchmark signals used in this study. 
For each signal type, the table reports the number of qubits $n$, the selected Fourier truncation parameter $m$, and the resulting CNOT count, circuit depth, and TD obtained from Qiskit simulations using the method of~\cite{moosa2023linear}.}
\label{tab:FSL}
\setlength{\tabcolsep}{4pt}
\renewcommand{\arraystretch}{1.1}
\begin{tabular}{c c c c cc cc cc c}
\toprule
factor &Periodic & Piecewise & Sinc & Single Gaussian & Multi Gaussian & PPG \\
\midrule
$n$               & 8      & 10     & 15     & 15     &   15   & 16 \\
$m$               & 7      & 7      & 6      &  5     &   6    & 12 \\
CNOT              & 576    & 615    & 491    & 364    & 491    & 17k \\
Depth             & 1028   & 1044   & 584    & 338    & 584    & 33k \\
TD                & 0.0377 & 0.0869 & 0.0235 & 0.0089 & 0.0104 & 0.0692 \\
\bottomrule
\end{tabular}
\end{table*}
This work introduced a simple and effective classical–to–quantum embedding strategy for efficiently loading compressible data into quantum registers.
The proposed hybrid method offers lower gate count and circuit depth, making it a practical strategy for data loading on near-term devices. The simulation results presented in section~\ref{sec:results} show that the hybrid approach substantially outperforms EAE in two key metrics, CNOT count and circuit depth, while maintaining high fidelity.

Beyond the comparison with EAE, it is also important to situate our approach within the broader family of approximate QSP techniques 
designed to reduce the cost of preparation. These include the FSL~\cite{moosa2023linear}, the Walsh Series Loader
(WSL)~\cite{zylberman2024efficient}, and QSVT-based schemes~\cite{mcardle2022quantum, o2025quantum, rosenkranz2025quantum}. 
Since our approach is fully ancilla-free, the most meaningful comparison is with the FSL, which also prepares states without additional qubits. 
In contrast, the WSL, which even offers shallower quantum circuits than FSL and block-encoding-based schemes, rely on ancillary qubits, making a direct comparison of circuit width and overall resource cost less appropriate for our setting.

To enable a fair evaluation, we applied the FSL to the same benchmark signals used in our hybrid framework. 
The FSL approximates a signal via a truncated Fourier series with $2^{m+1}$ modes, encoded in an $(m+1)$-qubit register. Loading these coefficients requires a unitary whose gate complexity and depth scale as $\mathcal{O}(2^m)$, followed by entangling layers and an inverse QFT over all $n$ qubits. 

Table~\ref{tab:FSL} summarizes the required quantum resources, where $m$ is chosen so that the final fidelity is comparable to the results of our hybrid method. 
For the lossless cases, periodic and piecewise-constant signals, FSL does not reach the fidelity obtained by our hybrid approach, even by increasing the FSL mode ($m$) and requiring $\sim 7\times$ and $\sim 5\times$ more CNOT gates, respectively. This highlights the advantage of transform-domain sparsity when the underlying representation is exactly sparse. 

For approximate QSP on smooth synthetic signals (sinc, Gaussian, and Gaussian mixture), the trend reverses. When matched at comparable TDs, the FSL achieves lower gate counts and shallower depths than the hybrid approach.
This behavior reflects the different ways each method trades accuracy for circuit complexity. 
Improving the accuracy of the FSL requires increasing $m$, affecting only a cost term scaling as $\mathcal{O}(2^{m})$. 
In contrast, improving accuracy in our framework requires adjusting the decomposition level $L$ and lowering the threshold $\tau$, 
which increases the number of retained coefficients and thus the cost of the SQSP stage, scaling as $\mathcal{O}(\mathrm{poly}(n))$. 
For the parameter regimes studied here ($n = 15$ with $m = 5$ or $6$), this polynomial term becomes larger than the corresponding $\mathcal{O}(2^{m})$ contribution in the FSL, making it the dominant overhead.

The most distinctive case is the PPG signal.
For equal fidelity, the FSL requires a comparable number of CNOT gates but produces circuits that are approximately $32\%$ deeper than those generated by the hybrid method. This contrast arises from the fundamentally different loading mechanisms. FSL must encode the entire signal using a global sinusoidal basis. At the same time, our hybrid method loads only the significant wavelet-domain coefficients via SQSP, where the localized structure and inherent denoising of wavelet transforms provide a much more efficient representation for real PPG signals.

The relative performance trends are consistent with the known error behavior of the FSL. As shown in~\cite{moosa2023linear}, the FSL achieves exponentially decaying error in the truncation parameter $m$, with the rate determined by the smoothness of the target function. This explains its strong performance on smooth synthetic signals. However, these guarantees do not transfer to irregular, noisy, or nonstationary data such as PPG recordings. In such settings, global Fourier approximations are less effective, while our transform-based sparsification remains robust

Taken together, these findings demonstrate that classical sparsification, combined with quantum decompression, provides a viable and flexible route to high-dimensional QSP. While the approach is not universal, its performance depends on the compressibility of the input; it aligns closely with the structural redundancy found in most real-world classical data and therefore constitutes a promising and scalable strategy for near-term quantum devices.

\section{Conclusion}\label{sec:conclusion}
We have presented a transform-agnostic, ancilla-free hybrid classical–quantum strategy that reduces the exponential overhead of EAE to a polynomial-cost procedure for a broad class of compressible data. 
Numerical simulations across synthetic and real biomedical signals demonstrate that classical sparsification can be directly translated into substantial reductions in CNOT count and circuit depth, while maintaining high reconstruction accuracy.
In particular, the hybrid method matches or closely approaches the performance of state-of-the-art ancilla-free approximate QSP techniques on synthetic signals, and exhibits clear advantages on non-stationary real-world data where global Fourier-based loaders need a deeper quantum circuit.

The present implementation, based on the DHWT and simple fixed-threshold pruning, should be viewed as a baseline instance of a more general hybrid framework.
Future work can proceed in two directions.
First, develop more efficient strategies for compressed data loading on the quantum system. This could include a novel SQSP method or using an ancillary qubit to reduce the polynomial overhead.
Second, evaluating the framework on broader data modalities—particularly those with established classical compression algorithms would help clarify its generality and practical limits.

Overall, our results indicate that combining classical sparsification with quantum decompression offers a flexible and scalable route to high-dimensional state preparation, making hybrid QSP a promising tool for near-term quantum applications.

\section*{Acknowledgments}
The authors would like to thank Magdalena Stobińska for her support and guidance during this work.
This research was supported by the National Science Centre, Poland, under the ‘Sonata Bis’ Project No.~2019/34/E/ST2/00273.

\bibliographystyle{unsrt}
\bibliography{mybibliography}
\end{document}